\documentstyle[prl,epsf,aps]{revtex}
\twocolumn
\begin{document}
\draft
\title{Giant negative magnetoresistance in semiconductors
doped by multiply charged deep impurities}
\author{ S.D. Ganichev\cite{sdg}, H. Ketterl,  W. Prettl}

\address{  Institut f\"ur Exp. und Angew. Physik,
Universit\"at Regensburg, 93040 Regensburg, Germany}

\author{I.A. Merkulov, V.I. Perel, I.N. Yassievich, A.V. Malyshev}
\address{A.F. Ioffe Physicotechnical Institute, RAS, St. Petersburg,
194021, Russia}

\date{\today}

\maketitle
\begin{abstract}
A giant negative magnetoresistance has been observed in
bulk germanium doped with multiply charged deep impurities.
Applying a magnetic field the resistance may
decrease exponentially  at any orientation of the
field.
A drop of the resistance as much as about 10000\% has
been measured at 6~T. The effect is attributed to the spin
splitting of impurity ground state with a very large
g-factor in the order of several tens depending on impurity.
\end{abstract}



It is surprising that in well investigated transport
properties of bulk semiconductors, particularly in the best
known material germanium,
until now new and previously not observed phenomena can be found.
Here we report on a giant negative magnetoresistance in Ge which shows
sizable effects already at very small magnetic field
strengths. An exponential drop of the resistance with rising
magnetic field, which may be more than two orders of
magnitudes,
occurs in a {\em parallel} as well in perpendicular  orientation of current
and magnetic field.

Negative magnetoresistance has attracted much interest in
the last decades due to the large variety of physical
phenomena causing a drop of the resistance of semiconductors
in an external magnetic fields.
One of the  striking effects is the low
temperature giant negative magnetoresistance observed in
disordered structures in
magnetic fields with a variable range  hoping
regime due to quantum interference leading
to weak
localization~\cite{Shklovskii84,Shklovskii85,Faran,Schirmacher,Shklovskii90,Mares}.
Other
important mechanisms of giant
negative magnetoresistance in semiconductors are magnetic field controlled
metal-insulator transitions\cite{Hellman},
removal of a minigap in a semiconductor
superlattice\cite{Lakrimi}, and magnetic field suppression of
spin-disorder scattering~\cite{Awschalom,Prinz}.
The application of a magnetic field on magnetic
perovskites aligns the spins in
different magnetic domains thereby lowering
the energy barrier for carriers and yielding a colossal negative
magnetoresistance\cite{Wagner}.
A negative magnetoresistance occurs also
in carbon nanotubes which has been shown to exhibit ballistic electron
transport\cite{Frank}, the increase of conductivity
has been attributed to a
magnetic field induced increase of the density of states in the vicinity of
the Fermi level\cite{Lee}.
The giant negative magnetoresistance reported here has only been
observed in samples doped with multiply charged impurities and
could not be detected in materials with only singly charged
impurities.

The experiments have been carried out on Ge:Hg, Ge:Cu, and
Ge:Ga. In germanium Hg and Cu are deep acceptors doubly and
and triply charged, respectively, whereas Ga is a singly
charged shallow acceptor. The binding energies of holes on
Hg are 90~meV and 230~meV for detachment of the first and the second
hole, respectively. From Cu three holes may be removed with
the binding energies 40~meV, 320~meV, and ($E_g - 260$)~meV
where $E_g$ is the energy gap.
The
hydrogen-like shallow impurity Ga has an ionization energy
of about 10~meV. The doping levels were in the range from
10$^{14}$ to 3x$10^{15}$~cm$^{-3}$. The typical size of the samples was
5~x~3~x~1~mm$^3$. One pair of ohmic contacts
were prepared on opposite faces. The samples were fixed in a temperature
variable cryostat. The resistance of the samples in the dark
has been obtained from the low voltage ohmic range of
current-voltage characteristics.
A magnetic field $B$ up to
6~T could be applied parallel and perpendicular to the
current flow by a superconducting magnet.

\begin{figure}
   \centerline{\epsfxsize 8cm \epsfbox{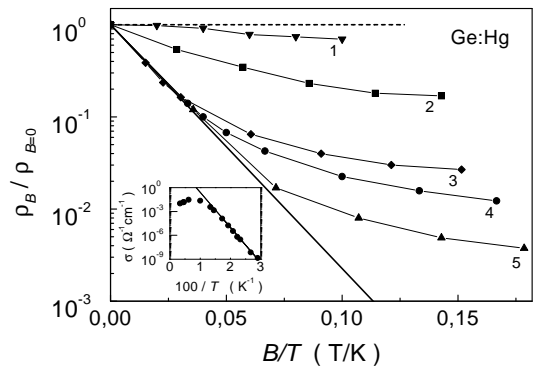}}
\caption{ A log-lin plot of the magnetoresistance
$\rho_B/\rho_{B=0}$ of Ge:Hg as a function
of the magnetic field strength $B$ normalized by the
temperature $T$ in the range $B = 0\dots 6$~T and for various
temperatures: 1- 55~K, 2- 40~K, 3- 38~K, 4- 35~K, 5- 33~K.
The full is a fit to $\exp(a B/k_{B}T)$ with $a =
5.8$~meV/T. The inset shows an Arhenius plot of the
conductivity at zero $B$.}
\label{F1}
\end{figure}

The conductance, $\sigma = 1/\rho$, where $\rho$ is the sample resistivity,
measured at zero magnetic field is shown as a function of
the inverse temperature, $1/T$, is plotted in the insets of
Fig.~1 and~2 for Ge:Hg and Ge:Cu, respectively. At low
temperatures the
temperature dependencies exhibit a clear Arhenius behaviour
determined by the corresponding binding
energies. All magnetoresistance measurements have been
carried out in these temperature ranges.

\begin{figure}[t]
   \centerline{\epsfxsize 8cm \epsfbox{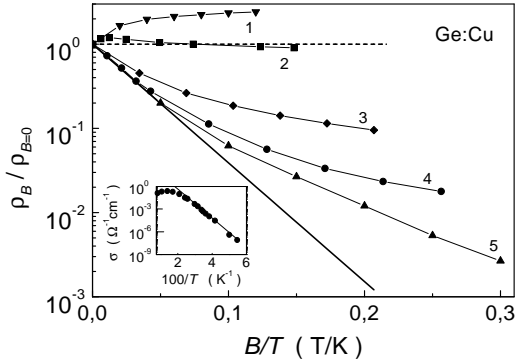}}
\caption{A log-lin plot of the magnetoresistance
$\rho_B/\rho_{B=0}$ of Ge:Cu as a function
of the magnetic field strength $B$ normalized by the
temperature $T$ in the range $B = 0\dots 6$~T and for various
temperatures: 1- 50~K, 2- 40~K, 3- 29~K, 4- 25~K, 5- 20~K.
The full is a fit to $\exp(a B/k_{B}T)$ with $a =
2.8$~meV/T. The inset shows an Arhenius plot of the
conductivity at zero $B$.}
\label{F2}
\end{figure}

In Fig.~1 the resistance of a Ge:Hg sample is shown as a
function of the magnetic field~$B$ normalized by the
temperature~$T$ for various, but for each measurement
constant temperatures. At low temperatures (curves~5,~4,
and~3) and small magnetic field strengths ($\sim 2$\,T) the
resistance drops exponentially with the same slope for
different temperatures. At higher field strength the
resistance saturates. At higher temperatures (curves~1 and~2
in Fig.~1) the magnetic field dependence gets weaker and
finally the negative magnetoresistance changes to positive
magnetoresistance. In the case of the perpendicular
geometry, the negative magnetoresistance is still present at
low temperatures but it is substantially smaller than in the
parallel geometry. This is caused by a compensation due to
the ordinary positive magnetoresistance in transverse
magnetic fields.

The analogous measurements on Ge:Cu are shown in Fig.~2. The
results are qualitatively the same with the difference that
the slope is here only one third of that of Ge:Hg.

The strength of the negative magnetoresistance is
independent on compensation ratio  in the
investigated range  $N_D / N_A = 0.18$ to $0.6$ at
low temperatures but gets dependent at higher temperatures
where a substantial free carrier density exists in the band.
This is shown in Fig.~3 where the resistance as a function
of~$B/T$ at constant~$T$ for various temperatures and for two
compensation ratios is plotted. The inset show the Arhenius
plot of the conductivity.

\begin{figure}[t]
   \centerline{\epsfxsize 8cm \epsfbox{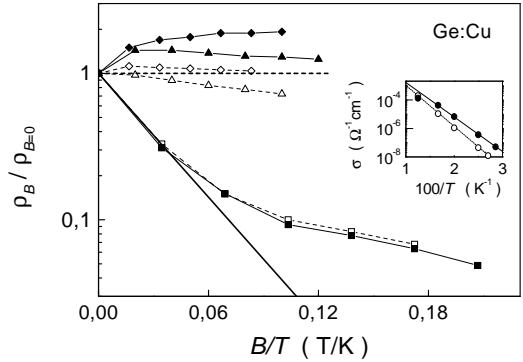}}
\caption{A log-lin plot of the magnetoresistance
$\rho_B/\rho_{B=0}$ of Ge:Cu as a function
of the magnetic field strength $B$ normalized by the
temperature $T$ in the range $B = 0\dots 6$~T and for various
temperatures and for two compensation ratios.
Diamonds, triangles, and squares correspond to $T= 60$~K, 50~K and 29~K,
respectively. Full symbols: $N_A = 1\cdot 10^{15}$~cm$^{-3}$, $N_D/N_A = 0.18$;
open symbols: $N_A = 3\cdot 10^{15}$~cm$^{-3}$, $N_D /N_A = 0.6$.
The full is a fit to $\exp(a B/k_{B}T)$ with $a =
2.8$~meV/T.
The inset shows an Arrhenius plot of the conductivity at zero $B$ for both
materials.}
\label{F3}
\end{figure}

The negative magnetoresistance has only been observed in the
dark and in a temperature range where only a small fraction
of the impurities were ionized. If the samples were
irradiated by visible or infrared light with photon energies
larger than the impurity binding energies, the negative
photoconductivity vanished. In the case of positive
magnetoresistance (at high temperatures) irradiation did not
affect the resistance ratio $\rho_B/\rho_{B=0}$.

With the singly charged shallow acceptor Ga in germanium
only positive magnetoresistance could be detected down to
liquid helium temperature.

The observations that a giant negative magnetoresistance
occurs only in materials doped with multiply charged impurities
and that the resistance decreases exponentially with rising
magnetic field in a significant range of temperature and
magnetic field strength give a key for a qualitative
understanding of the phenomenon. The exponential drop of the
resistance indicates a decrease of the impurity binding
energy being linear as a function of the magnetic field. The
different behaviour of singly and doubly charged impurities
showing positive and negative magnetoresistance,
respectively, will be discussed on the basis of a comparison
with magnetic field dependence of the ionization energy of
neutral hydrogen and helium atoms.
In both cases the low energy edge of the continuum states
does not depend on magnetic field because the Landau
diamagnetism ($\Delta \varepsilon_L= \Delta \varepsilon = \hbar \omega_c
/2$) is
compensated by the Pauli spin paramagnetism ($\Delta
\varepsilon_P = - \Delta \varepsilon= -\mu_B B = -\hbar\omega_c /2$).
Here $\mu_B$ and $\omega_c$ are the Bohr magneton and the
cyclotron frequency, respectively.
For hydrogen atoms in relatively low magnetic fields the
energy of the ground state level, $E_H(B)$, goes down due to spin
paramagnetism.  The diamagnetic contribution is
vanishingly small. Thus, the ionization energy of hydrogen
atoms linearly increases with rising magnetic field. For
helium atoms the situation is just the other way round, the
binding energy decreases. The reason is that there are now
two electrons with zero total spin
on the $1s$
shell.
Hence,
the energy of this pair of electrons is independent of
magnetic field strength. After ionization of the first electron,
one electron remains on the shell whose ground state energy  level
goes down in the same way like that of the H-~atom. Thus, the
ionization energy of the first electron, $E_{He^0}$,
decreases by the value of $\Delta E_{i1}$ and that of the second electron,
$E_{He^+}$,
increases by the value of $\Delta E_{i2}$
as a function of the magnetic field. Therefore, $\Delta
E_{i1} = -\Delta E_{i2} = -\hbar\omega_c /2$. The scheme of
the energy levels involved in this discussion is sketched in
Fig.~4.

\begin{figure}
   \centerline{\epsfxsize 8cm \epsfbox{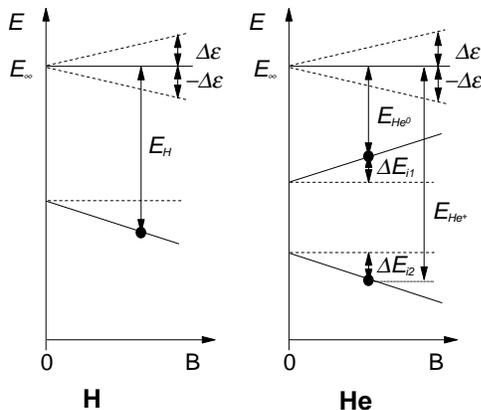}}
\caption{Scheme of the energy levels of  H-like and He-like atoms in
magnetic field. $E_A$ is the continuum edge, $E_H(B)$, $E_{He^0}(B)$, and
$E_{He^+}(B)$ are ionization energies. $\Delta E_{ij}$ $(j=1,2)$  are
changes of ionization energies in a magnetic field. $\Delta \varepsilon$ is
the magnitude of paramagnetic and diamagnetic shifts of the continuum edge
compensating each other.}
\label{F4}
\end{figure}

This simple picture seems to apply to singly and multiply
charged acceptors in semiconductors with rather complicated
valence band where free holes are characterized by spin~$3/2$.
For  singly charged impurities, as in
the case of hydrogen atoms, the ionization energy increases
with rising magnetic field due to spin paramagnetism of
holes in the ground state. Thus with rising magnetic field
strength the density of free holes decreases yielding a
freezing--out of free carriers.
This is in good
agreement to the positive magnetoresistance observed in
Ge:Ga.
For centers with two holes (doubly charged impurities)
the ionization energy of the
two--particle ground state can decrease linearly with the magnetic
field as in the case of atomic helium.
In contrast to the atomic situation, however,
the edge of the continuum is not independent of the magnetic field strength
and increases the binding energy~\cite{Hensel}.
Thus, in order to obtain the
decrease of the binding energy which results in the observed
exponential
increase of
free carrier concentration, the paramagnetic shift of the second hole
$|\Delta E_{i2}|$ must override
the shift of the bottom of the valence band $\Delta E_0$ and
the ground state shift of the acceptor with two holes.

The analysis of the measurements in the range of exponential
decrease of the resistance (low temperatures and $B$= 0 - 2~T)
using $\Delta E_{i1} = -g \mu_B B$, where $\Delta E_{i1}$ is the
change of the impurity binding energy,
leads to a g-factor $g = 100$ for Ge:Hg and
$g = 48$ in the case of Ge:Cu (note Ge:Cu is triply charged).
Therefore the effect of such a large g-factor overrides any
shift of the band edge and the ground state of acceptor with two holes
in the magnetic field.

The origin for such giant g-value remains unclear.
Calculations based on
the effective-mass approximation after~\cite{Merkulov1,Merkulov2}
yield a ground state g-factor  varying from about $-1$ for the
shallow level ($E_A \ll \Delta_{so}$) to about $10$ for deep
centers
($E_A \sim \Delta_{so}$). Here $E_A$ and $\Delta_{so}$ are
the acceptor ground state energy and
spin-orbit energy splitting, respectively.
These  theoretical estimations show that the g-value
increases with the ground state energy, which is
qualitatively in agreement with the
experimental data.

The experimentally observed deviation from the exponential
drop of the resistance at high magnetic fields and
intermediate temperatures (Figs.~1 and~2) is due to a large
increase of free carrier concentration which show a positive
magnetoresistance. The same effect of free carriers causes
the influence of compensation ratio on the magnetoresistance
(Fig.~3).

In summary, in contrast to all established mechanisms
of negative magnetoresistance,
the giant negative magnetoresistance experimentally observed
in germanium is due to a large shift of the thermal
population of the band in a magnetic field. The exponential
decrease of resistance requires a linear splitting of the
impurity ground state in the magnetic field with an
astonishingly large g-factor. The large magnitude of
g-factor needs further investigation in order to explain it.

Financial support by the DFG, the RFFI and the NATO linkage program
are gratefully acknowledged. The authors would like to thank M. Kagan
for helpful discussions.

\end{document}